\documentclass[journal]{IEEEtran}
\usepackage[T1]{fontenc}
\usepackage[latin9]{inputenc}
\usepackage{amsmath}
\usepackage{amssymb}
\usepackage{amsthm,bm}
\usepackage{url}
\usepackage{cite}
\usepackage{babel,verbatim,balance}
\usepackage[rgb]{xcolor}
\usepackage{graphics, graphicx}
\usepackage{acronym}
\usepackage{upgreek}
\usepackage[bookmarks,colorlinks]{hyperref}
\usepackage{setspace,gensymb}

\begin{document}

\title{Self-Adaptive RISs Beyond Free Space:\\ Convergence of Localization, Sensing and Communication under Rich-Scattering Conditions}

\author{Chlo\'e Saigre-Tardif and Philipp del Hougne
\thanks{Chlo\'e~Saigre-Tardif and Philipp~del~Hougne are with Univ Rennes, CNRS, IETR - UMR 6164, F-35000, Rennes, France.
}
\thanks{\textit{Corresponding Author: Philipp del Hougne (e-mail: philipp.del-hougne@univ-rennes1.fr).}}
}

\maketitle
\begin{abstract}

We discuss the need for a confluence of localization, sensing and communications if RISs are to be deployed in a self-adaptive manner in the dynamically evolving rich-scattering settings that are typical for 6G deployment scenarios such as factories. We establish that in such problems the rich-scattering wireless channels are subject to a  highly nonlinear deterministic double-parametrization through both the RIS and uncontrolled moving objects. Therefore, acquiring full context-awareness through localization and sensing is a prerequisite for RIS-empowered communications. Yet, the byproducts of this daunting communications overhead can feed many appliances that require context awareness, such that overhead concerns may vanish. We illustrate the essential steps for operating a self-adaptive RIS under rich scattering based on a prototypical case study. We discover that self-adaptive RISs outperform context-ignorant RISs only below a certain noise threshold that depends, among other factors, on how strongly uncontrolled perturbers impact the wireless channel. We also discuss ensuing future research directions that will determine the conditions under which RISs may serve as technological enabler of 6G networks.
\end{abstract}

\section{Introduction}

While fifth generation (5G) wireless communication networks are still being rolled out, researchers already actively explore how the anticipated needs for sixth generation (6G) wireless communication networks can be met. New applications like virtual reality and the wireless networked control of robotic systems (smart factory, automated driving, etc.) are expected to require up to 1000-fold increases of peak data rates and at least 10-fold improvements of energy and spectral efficiencies~\cite{you2021towards}. One promising technological enabler for 6G may be reconfigurable intelligent surfaces (RISs) which endow wireless environments with programmability~\cite{di2019smart}. Thereby, the physical layer system design is no longer restricted to the radiated waveforms but additionally includes the wireless channels, enabling a clear paradigm shift ``beyond Shannon'' who's famous law assumes uncontrolled channels.

But the full potential of this ``smart radio environment'' concept can only be reaped in practice if the RIS can self-adapt its configuration to a dynamically evolving deployment setting. First concepts for self-adaptive RISs are being studied for deployment scenarios in free space~\cite{alexandropoulos2021hybrid}, that is, within an essentially empty and thus very simple radio environment. Corresponding experimental efforts are conducted in anechoic (echo-free) rooms~\cite{ma2019smart,ma2020smart}. In such free-space scenarios, the RIS-parametrized channel is a three-element linear cascade: \textit{i)} propagation from the base station (BS) to the RIS, \textit{ii)} modulation by the RIS, and \textit{iii)} propagation from the RIS to the user equipment (UE). Therefore, a free-space self-adaptive RIS must sound two channels (BS-RIS and RIS-UE) in order to update itself, which is recognized as implying substantial overhead that may even outweigh the benefits of RIS-parametrized channels.

However, many deployment scenarios of 6G are not compatible with this free-space hypothesis. 6G is expected to consist of all-spectra integrated networks, ranging from microwave systems that offer wide coverage at low cost, via millimeter wave systems, all the way to even terahertz or optical systems with unprecedented bandwidth~\cite{you2021towards}. At microwave frequencies, even an office room can give rise to substantial rich scattering that invalidates the free-space assumption, as seen in experiments at 2.45~GHz~\cite{del2019optimally}. Envisioned factory settings for Industry 4.0 are full of metallic elements that certainly strongly scatter microwaves and millimeter waves. Therefore, the analysis of self-adaptive RISs must be fundamentally rethought beyond the free-space hypothesis, and surprisingly optimistic conclusions about the associated overhead may follow.

A pivotal point is the qualitatively very different parametrization of rich-scattering wireless channels through RISs. As noted above, free-space channels are linearly parametrized by a RIS. In contrast, the RIS-parametrization of rich-scattering channels is highly nonlinear because any given ray can encounter multiple RIS elements as it ricochets on its way from BS to UE. A rich-scattering self-adaptive RIS must develop a detailed deterministic understanding of this nonlinear parametrization. Merely sounding BS-RIS and RIS-UE channels would not be of much help. Instead, a rich-scattering self-adaptive RIS must essentially gain complete context awareness (localize transceivers and scatterers, sense shapes and postures, etc.) and combine these insights with its understanding of the nonlinearly parametrized channel in order to self-adapt. The complexity of these localization and sensing tasks may appear daunting, especially with regard to overhead concerns.

Fortunately, the results from localization and sensing tasks that a rich-scattering self-adaptive RIS must perform as auxiliary step to serve its primary communication purpose are actually themselves highly sought after. Gaining context awareness is essential for emerging services in areas including smart health care, touchless human-machine interaction, autonomous mobility, and security. The potential of RISs for localization~\cite{wymeersch2020radio,del2021deeply} and sensing~\cite{saigre2022intelligent} has not gone unnoticed, but these explorations are so far decoupled from communications-related RIS deployments. Yet, outside the specific RIS context, there is already a general tendency toward Integrated Sensing and Communication (ISAC)~\cite{ma2020joint,de2021convergent,zhang2022holographic} that is fueled by striking similarities between sensing and communication systems, and motivated by limited spectral resources and power consumption constraints. Of course, this ISAC trend is expected to extend to RIS-empowered systems. 
But given the above analysis, context-aware services can be fed with localization and sensing information as a byproduct of a self-adaptive RIS-empowered communication system under rich-scattering conditions, implying virtually zero overhead. In RIS-empowered rich-scattering settings, sensing is thus a prerequiste for communication, in contrast to typical ISAC concepts that seek to unify two systems serving different purposes, one for sensing and one for communication.

In this paper, we introduce the concept of self-adaptive RISs beyond the oftentimes questionable free-space assumption. We begin by reviewing the usually overlooked nonlinearity of the parametrization of rich-scattering wireless channels. Then, with the help of an illustrative case study, we introduce the essential ingredients of operating a self-adaptive RIS under rich-scattering conditions. We explain how its overhead automatically provides full context awareness. We conclude by looking forward to key research challenges and open questions that will determine the feasibility of deploying RISs in 6G networks under realistic conditions.

\section{Parametrized Rich-Scattering Channels}
\label{sec_param}

\begin{figure}
\centering
\includegraphics [width = \linewidth]{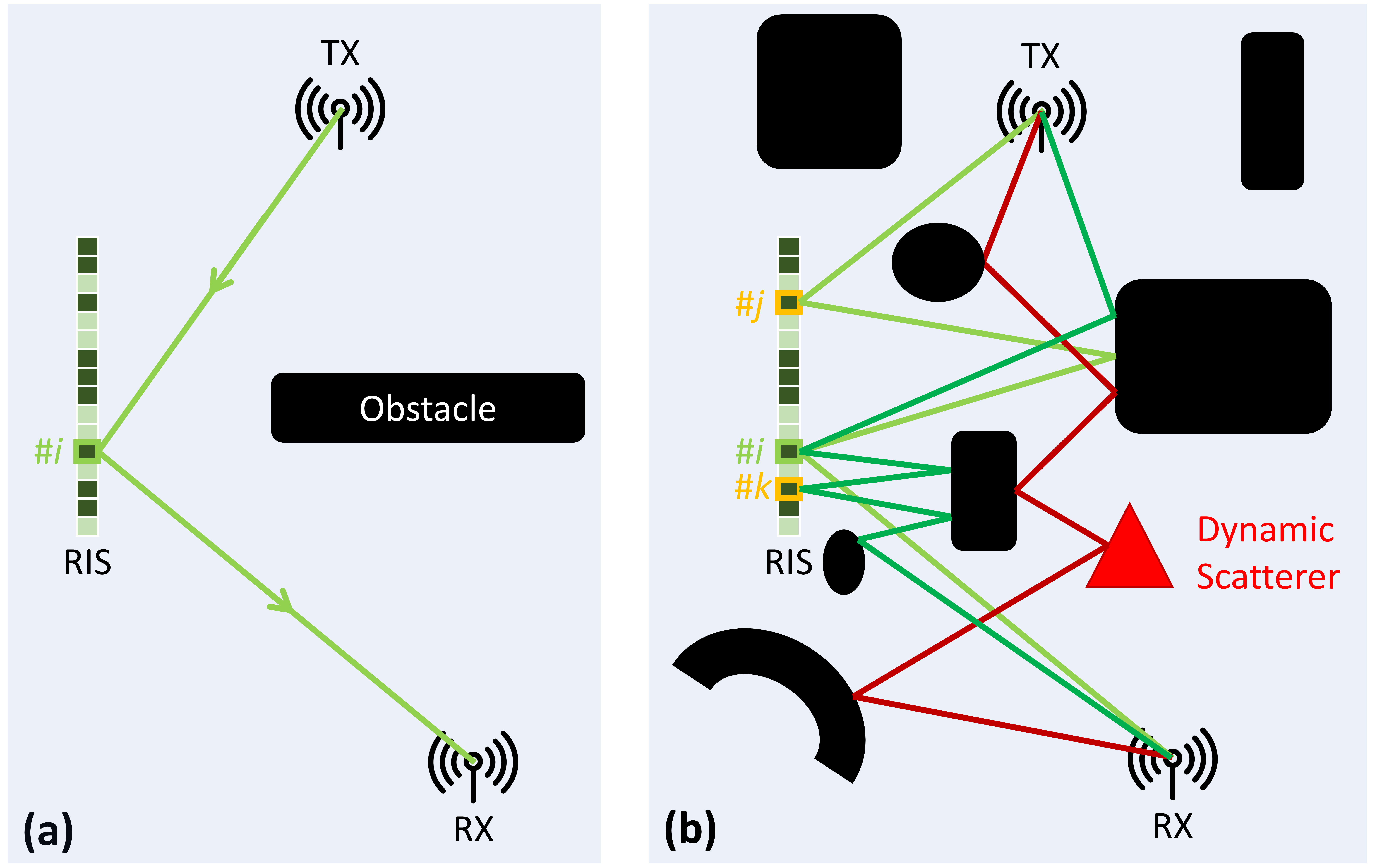}
\caption{(a) A quasi free-space wireless channel is linearly parametrized by a RIS. (b) A fading rich-scattering wireless channel is nonlinearly double-parametrized by a RIS and moving perturbers.}
\label{fig1}
\end{figure}

Wireless communications has to date primarily been a model-driven field. The emerging RIS paradigm now introduces a \textit{deterministic} parametrization of wireless channels that is fundamentally incompatible with conventional \textit{statistical} models of fading channels. Current attempts at marrying together these two worlds take the free-space linear RIS-parametrization from Fig.~\ref{fig1}a as starting point to imagine a linear cascade of fading, interaction with the RIS, and further fading. Despite the appealing simplicity of this approach, it is unsuitable for analysing RIS-parametrized rich-scattering environments because it violates fundamental (wave) physics. First, it fails to capture mesoscopic correlations, originating from ricocheting rays, that yield a highly nonlinear RIS-parametrization of the wireless channel. As sketched in Fig.~\ref{fig1}b, rays connecting a transmitter to a receiver via the $i$th RIS elements are typically also bouncing off other RIS elements, such that the wireless channel cannot be formulated as a linear function of individual RIS element configurations. Second, the use of random matrices to emulate fading violates causality because random matrices have no notion of space and hence cannot guarantee that a signal is received after (and not before) it is emitted.

It is important to not confuse the uncontested linearity of the wave equation with the nonlinear parametrization of the wireless channel. The former refers to the superposition property, namely that the output corresponding to multiple simultaneous inputs is a linear combination of the outputs that would be obtained if each input was injected individually. The latter refers to how the channel (rather than the output) depends on the boundary conditions (rather than the input). 
The degree of nonlinearity of a wireless channel's RIS-parametrization depends on the amount of reverberation in the environment. This is intuitively obvious: if an environment presents less attenuation, the lifetime of rays is longer such that they can encounter more RIS elements during their lifetime. In the limiting case of free space without any reverberation, the linear RIS-parametrization is recovered. In enclosed metallic environments such as inside vessels or factories, the attenuation is weaker than in partially open and lossy environments like office rooms. 

From the electromagnetic perspective, RISs can be understood as arrays of \textit{purposefully controlled} perturbers of the wireless channel. However, fading rich-scattering environments are additionally parametrized by \textit{uncontrolled} perturbers whose locations and shapes dynamically evolve. An example would be the body of a human who moves inside an office room and assumes various postures (upright, sitting, etc.). The wireless channel is therefore subject to a \textit{deterministic non-linear double-parametrization} through both the RIS configuration and a collection of parameters that describes locations and shapes of the uncontrolled perturbers. In order to determine a suitable RIS configuration that supports a desired communications objective one must \textit{i)} deterministically understand this complex parametrization, and \textit{ii)} estimate the current perturber parameters. The latter implies the acquisition of complete context awareness.

\section{Case Study}

\subsection{Scenario}

\begin{figure}
\centering
\includegraphics [width = \linewidth]{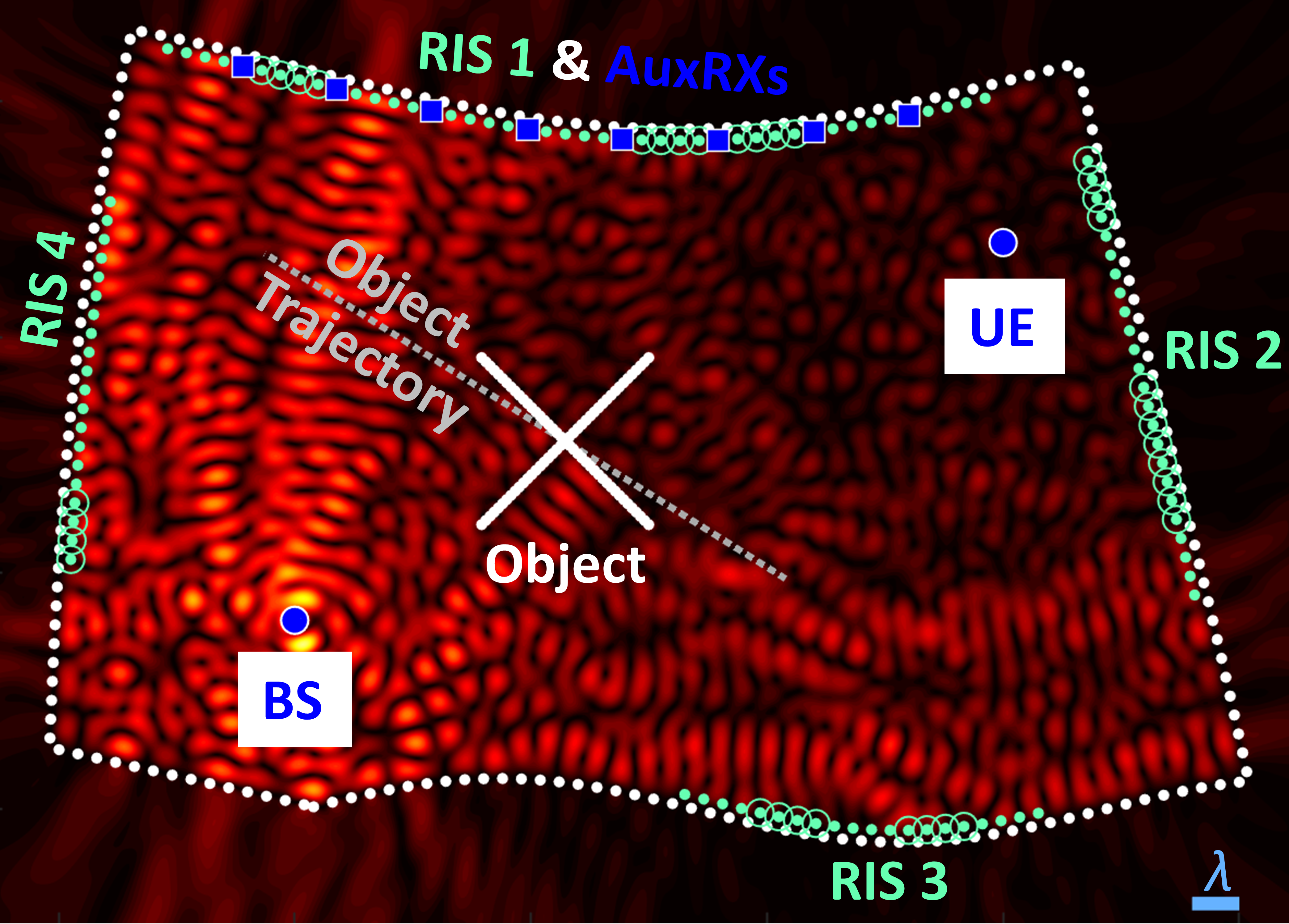}
\caption{Considered rich-scattering scenario in PhysFad~\cite{faqiri2022physfad}, comprising a many-wavelengths-large irregularly-shaped reverberant room (see wavelength $\lambda$ scalebar), a base station (BS), a user equipment (UE), a conformal four-part distributed RIS with eight auxiliary receivers (AuxRXs), and an object moving along the indicated trajectory. In addition, the object can assume three different shapes: cross (shown), circle, or square. The RIS consists of 25 1-bit programmable macro-pixels, each grouping together four adjacent elements (encircled elements are ON resonance, the others OFF resonance). The sketch is overlayed on the corresponding field magnitude map.}
\label{fig2}
\end{figure}

To demonstrate the essential aspects of operating a self-adaptive RIS under rich-scattering conditions, we present a case study of the prototypical example sketched in Fig.~\ref{fig2}: an enclosed indoor environment in which an object moves and changes shape (representing, for instance, a human). The uncontrolled perturbations are thus parametrized through: 

\begin{enumerate}
   \item Object location along its allowed trajectory.
   \item Object shape among its three allowed shapes.
 \end{enumerate}

\noindent In this example, the first item is a continuous parameter whereas the second one is a discrete parameter. 

\begin{figure}
\centering
\includegraphics [width = \linewidth]{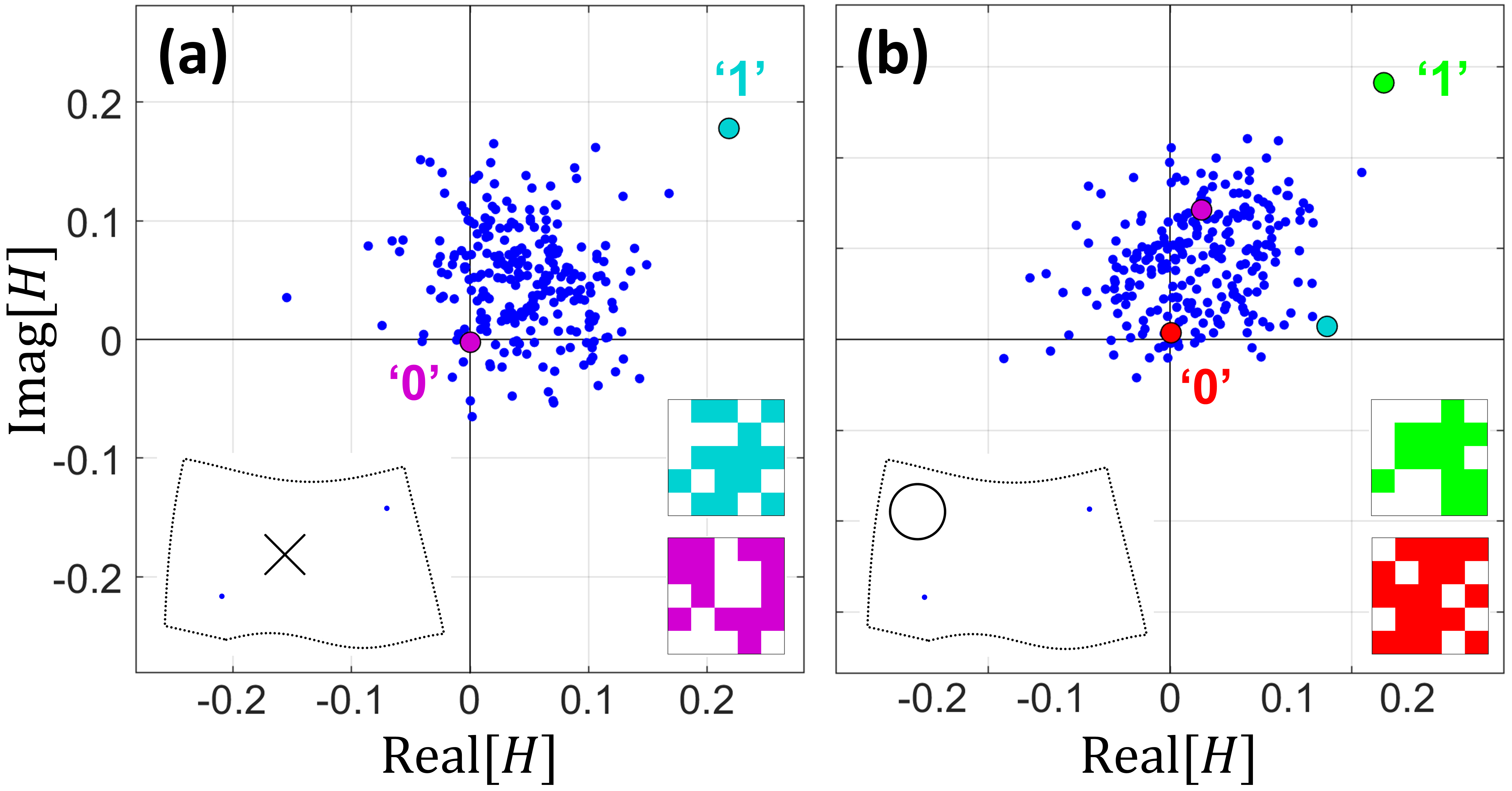}
\caption{RIS-empowered BASK in double-parametrized wireless channel $H$. For two different choices of perturber parameters (see bottom left insets), (a) and (b) show the wireless channel for 100 random RIS configurations (blue), as well as for configurations optimized to minimize (`0') or maximize (`1') the RSSI. The optimized configurations are displayed as inset. The wireless channels for the optimized configurations from (a) in the setting of (b) are also shown. A dot's distance from the origin is the RSSI.}
\label{fig3}
\end{figure}

The rich scattering is evident upon inspection of the field map in Fig.~\ref{fig2}. The isotropic field maxima, especially in the vicinity of the UE, evidence that rays from all possible directions are incident -- in sharp contrast to beam-forming concepts known from free space. In our case study, the RIS' desired communications functionality is binary amplitude-shift-keying (BASK) backscatter communication: the RIS is used as massive backscatter array to maximize (`1') or minimize (`0') the RSSI~\cite{zhao2020metasurface}. Relative to the average RSSI, maximization yields an improvement of around 10~dB whereas minimization achieves an RSSI close to zero -- see Fig.~\ref{fig3}a. But the necessary RIS configurations for this purpose strongly depend on the perturber parameters. Indeed, the two optimal configurations for the perturber parameters in Fig.~\ref{fig3}a are impossible to distinguish in terms of the corresponding RSSI for another combination of perturber parameters, as seen in Fig.~\ref{fig3}b. The strong dependence of the wireless channel on \textit{both} the RIS configuration and the perturber parameters is thus apparent in Fig.~\ref{fig3}. The self-adaptive RIS must hence perform both localization and sensing tasks to estimate the perturber parameters as prerequisite in order to subsequently optimize its configuration for the desired communications functionality.

\subsection{Channel Model}

As explained in Sec.~\ref{sec_param}, commonly used cascaded linear channel models for RIS-controlled rich-scattering wireless environments violate causality and fail to capture pivotal mesoscopic correlations. To the best of our knowledge, to date, the only available physics-compliant channel model for RIS-parametrized rich-scattering is PhysFad~\cite{faqiri2022physfad}; PhysFad is derived from first principles based on a coupled-dipole formalism and describes all wireless entities as dipoles or collections of dipoles with suitable properties (resonance frequency, absorption, etc.). As seen in Fig.~\ref{fig2}, continuous surfaces of scattering objects are discretized and represented as dipole fences. As detailed in Ref.~\cite{faqiri2022physfad}, the configuration of a RIS element is controlled via the resonance frequency of the corresponding dipole. The underlying Lorentzian resonator model automatically accounts for frequency selectivity and the intertwinement between phase and amplitude response. For simplicity, and in line with most existing experimental prototypes (e.g., Refs.~\cite{del2019optimally,del2021deeply}), we limit ourselves to 1-bit programmable RIS elements. 

In a given realization of the RIS-parametrized rich-scattering environment, the wireless channel depends in a non-trivial manner on the locations and properties of all dipoles needed to describe the considered scenario. The self-consistent PhysFad formalism involves a matrix inversion to correctly capture all mesoscopic correlations. Mathematical details are explained in Ref.~\cite{faqiri2022physfad} and not repeated here for the sake of brevity. Overall, PhysFad outputs the wireless channel $H$ between the designated transmitting and receiving dipoles for the specified double-parametrization (RIS and perturbers) of the considered wireless environment. We focus on the RSSI, that is $|H|$ assuming unity transmit power in our scenario. Throughout this paper, $H$ is evaluated at a single working frequency.

\subsection{Operation Principle}

The operation principle of a rich-scattering self-adaptive RIS can be broken down into two steps: localization and sensing to gain complete context awareness, followed by the identification of suitable RIS configurations for the desired communications functionality. Both steps require an offline calibration to adjust to the specific wireless environment under consideration as well as an online operation during runtime that is repeated once per coherence time.

\textbf{Localization and Sensing.} We tackle localization and sensing in rich-scattering conditions under a unified conceptual and technical framework as inverse problem: which perturber parameters explain the available field measurements? This approach strongly differs from its free-space counterpart where RIS-empowered localization~\cite{wymeersch2020radio} and sensing~\cite{saigre2022intelligent} rely on ray-tracing techniques (angle of arrival, etc.) and the first Born approximation (absence of multiple scattering), respectively. These underlying assumptions are incompatible with rich-scattering conditions where waves are strongly ricocheting: rays from all possible angles are incident and any coherent wavefront gets completely scrambled. Nonetheless, the sought-after location and sensing information is encoded in the scrambled waves -- hence the inverse-problem perspective. The key insight is that the field measured at an arbitrary location in a rich-scattering environment is the superposition of all reflections, and reflections off the objects of interest depend on their location, shape and orientation. 

A series of field measurements can hence act as a wave fingerprint that enables the unambiguous identification of the perturber parameters. Thanks to the configurational diversity offered by the RIS, this series of measurements can be performed at a single frequency and with a single or a few node(s). To that end, we must learn to recognize the characteristic fluctuations of the wireless channel(s) over a fixed series of random RIS configurations that are specific to each value of the sensing parameter. To avoid that the self-adaptive RIS requires cooperation with the UE, we endow it with its own auxiliary receivers. Given the statistical uniformity of random wave fields, these auxiliary receivers (AuxRXs) could be placed anywhere, but for the practical convenience of compactness we integrate them into a part of the distributed RIS in our case study (see Fig.~\ref{fig2}), similar to the free-space self-adaptive RISs from Refs.~\cite{ma2020smart,alexandropoulos2021hybrid}.

The required number of the auxiliary measurements for precise localization and sensing depends on how much information about the perturber parameter of interest can be extracted per auxiliary measurement:

\begin{enumerate}
   \item \textit{Amount of reverberation.} The longer the waves reverberate before being attenuated, the more sensitive the field becomes to a given perturber parameter~\cite{del2021deeply}.
   \item \textit{Perturbation size.} The larger the perturbation is, the more strongly it affects the field.
   \item \textit{Signal-to-noise ratio (SNR).} The higher the SNR is, the more information can be extracted per measurement.
 \end{enumerate}

To solve said inverse problems, fully connected artificial neural networks (ANNs) can be trained as surrogate inverse models. We use the auxiliary field measurements as input to the ANN, and the output is either the prediction of a continuous variable (perturber location) or a classification (perturber shape). Fully connected ANN architectures are deliberately chosen because the sought-after information is presumable \textit{not} encoded in local features but long-range correlations~\cite{del2020robust,del2021deeply} -- in contrast to computer vision problems where convolutional architectures excel. Because we train one ANN per perturber parameter, each ANN should be insensitive to fluctuations of the other perturber parameter(s) so that it can robustly estimate the assigned parameter~\cite{del2020robust}. A training data set with sufficiently representative parameter combinations is required.

\textbf{Communication.} The second step confronts us with an inverse-design problem: which RIS configurations best satisfy the needs of the desired communications functionality (here BASK backscatter communication)? The answer strongly depends on both the current perturber parameters and the specific wireless environment. Therefore, both the acquisition of full context awareness (via online localization and sensing as discussed above) and an offline calibration to the specific environment are unavoidable prerequisites for RIS-assisted communications in dynamic rich-scattering conditions. Two options present themselves for the type of offline calibration:

\begin{enumerate}
   \item \textit{Open loop.} During the offline calibration, a surrogate forward model is trained. This model takes RIS configuration and perturber parameters as inputs to predict the wireless channel in the specific setting. During runtime, the self-adaptive RIS makes use of this model, in combination with the sensed perturber parameters, to identify its optimal configuration once per coherence time.
   \item \textit{Closed loop.} During the offline calibration, optimal RIS configurations are learned for each possible combination of perturber parameters. During runtime, based on the sensed pertuber parameters, the most appropriate configuration is chosen out of the pre-stored code book.
 \end{enumerate}
 
 Successfully learning a surrogate forward model can be extremely challenging for the highly nonlinear parametrization associated with strong reverberation. Moreover, the online optimization once per coherence time can be costly in terms of latency and computational burden. Especially when the perturber parameter space is of manageable size, the closed-loop approach is hence very attractive: the nonlinear channel parametrization is not learned explicitly and looking up a suitable code-book entry during runtime is very fast. 
 However, the closed-loop approach requires control over the RIS to measure the channels corresponding to specific RIS configurations during the calibration in order to optimize the RIS configurations. In contrast, the open-loop approach offers the possibility of updating the calibration through observations with arbitrary RIS configurations during runtime.
 
\begin{figure}
\centering
\includegraphics [width = \linewidth]{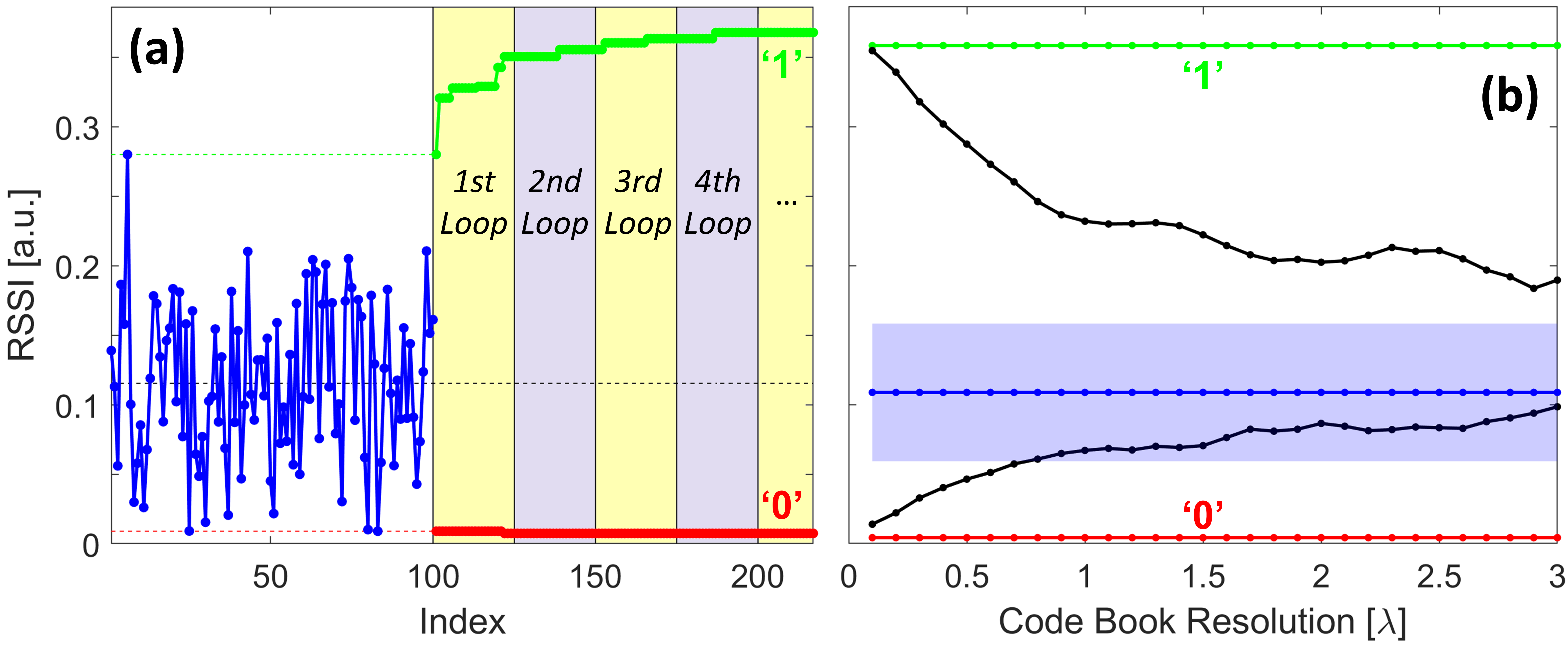}
\caption{Closed-loop self-adaptive rich-scattering RIS scheme for the backscatter BASK functionality. (a) Example iterative optimization of the RIS configuration for the parameter settings from Fig.~\ref{fig3}b. (b) Average performance dependence on the code-book resolution.  }
\label{fig4}
\end{figure}

The closed-loop approach must inevitably discretize continuous perturber parameters (in our example the object location) because the code book can only have a finite number of entries. For each discrete perturber parameter value, optimal RIS configurations for the desired communications functionality are identified during calibration. These optimizations are difficult due to the nonlinear parametrization and the 1-bit programmability constraint. An example of a simple iterative approach to such an optimization is shown in Fig.~\ref{fig4}a. First, the channel is measured for 100 random RIS configurations. Second, the best configuration is taken as starting point for an element-by-element iterative optimization. The latter must loop multiple times over each RIS element before convergence -- a clear signature of the nonlinear channel parametrization: the optimal configuration of the $i$th RIS element depends on the configuration of the other RIS elements. During runtime, the object will continuously vary its position and the localization step will predict a continuous position variable. The communication step will then look up the code-book entry that most closely corresponds to this predicted position. The expected performance loss due to the limited code-book resolution is plotted in Fig.~\ref{fig4}b. A sub-wavelength code-book resolution is clearly needed to avoid strong performance penalties. Thus, the closed-loop approach is suitable when the perturber parameter space is of a manageable size. Note that interpolating between code-book entries is not easily realizable due to the 1-bit programmability of the RIS elements.
 
\textbf{Simultaneous Localization, Sensing and Communication.} So far, we have discussed a self-adaptive RIS system in which communications are interrupted once per coherence time in order to perform localization and sensing with a dedicated series of RIS configurations, to subsequently update the RIS configurations used for communications. If the coherence time is long and if the RIS sequence for localization and sensing is short, then these interruptions are certainly tolerable. Such an ISAC system performs both localization and sensing as well as communications using the same hardware, and the context awareness obtained from localization and sensing is a prerequisite to successfully communicate. However, it would be even better if localization and sensing could be performed without dedicated RIS configurations but instead based on the RIS configurations used for communications. Thereby, interruptions of the communications to perform localization and sensing would be avoided.

The associated challenge lies in solving the inverse problem that underlies the localization and sensing. The previously discussed surrogate inverse model maps auxiliary channel measurements obtained for a \textit{fixed} series of RIS configurations to the sought-after perturber parameters. Given the strong channel dependence on the RIS configuration (see Fig.~\ref{fig3}), transposing this technique to auxiliary channel measurements obtained for an \textit{arbitrary} series of RIS configurations is very challenging. An open-loop approach is to train ANNs that take both the auxiliary measurements and the corresponding RIS configurations as input. A closed-loop approach suitable for manageable perturber parameter spaces is to train separate ANNs for each configuration from the communications code book. Even if the latter involves training a few hundred ANNs, it may still be an easier task than learning to be able to deal with any out of all possible $2^{25}$ configurations in the open-loop approach.

\subsection{Results}

Having established the operation principle of a self-adaptive RIS under rich-scattering conditions, we present indicative results for our prototypical scenario. Given the manageable perturber parameter space in our case study, we opt for the closed-loop approaches discussed above. In particular, thanks to the eight auxiliary receivers, sensing and localization are possible based on a single RIS configuration at the targeted 20~dB level of SNR in our problem. Hence, with each transmitted symbol the self-adaptive RIS can update its knowledge about the environmental context (object position and shape) and consequently the RIS configuration it uses for the BASK backscatter communications. In other words, localization and sensing are performed without dedicated RIS configurations that would interrupt communications.

\begin{figure}
\centering
\includegraphics [width = 0.6\linewidth]{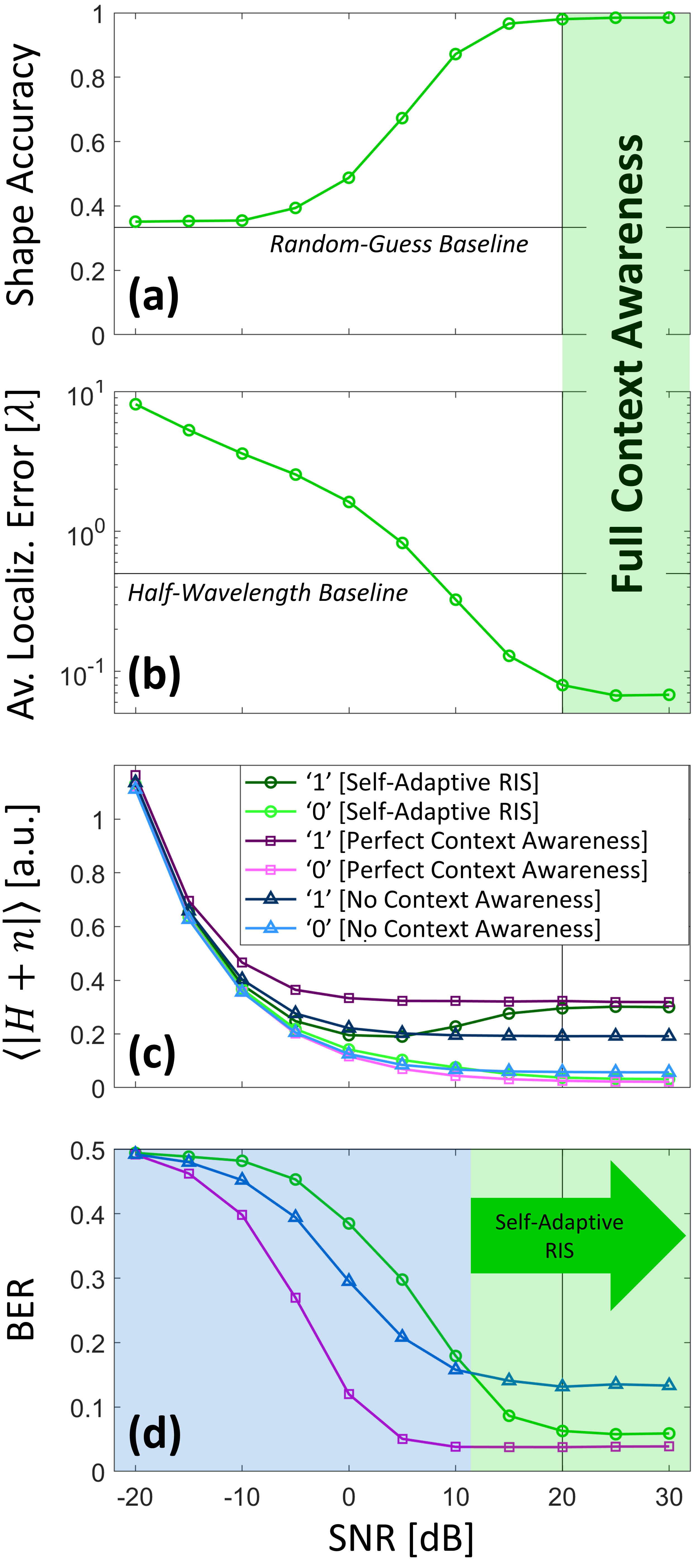}
\caption{ISAC with self-adaptive rich-scattering RIS performing BASK backscatter communication. (a) Object shape recognition accuracy. (b) Average object localization error. (c) Measured magnitude (signal plus noise) for the two possible BASK symbols (`0' and `1') for the discussed self-adaptive RIS (green) and two benchmarks: a RIS operated \textit{with perfect} context awareness (purple) and a RIS operated \textit{without any} context awareness (blue). (d) Corresponding BER curves for the three scenarios from (c).  }
\label{fig5}
\end{figure}

In Fig.~\ref{fig5}a, the accuracy with which the self-adaptive RIS recognizes the object's shape is plotted. For SNR values above the targeted 20~dB, the accuracy is close to unity; at low SNRs, it approaches the random-guess baseline. Similarly, in Fig.~\ref{fig5}b, the average error of the estimated object location is plotted in units of wavelength. Above the targeted SNR of 20~dB, the average error is on the order of a tenth of the wavelength. Note that this is well below the half-wavelength resolution baseline which is frequently assumed to be a fundamental bound -- however, there is no half-wavelength resolution limit because reverberation endows us with interferometric sensitivity and deeply subwavelength resolution under rich-scattering conditions~\cite{del2021deeply}. The first result is therefore that high-fidelity sensing and localization, equivalent to full context-awareness, are possible without dedicated RIS configurations, thus enabling uninterrupted communications. This real-time context awareness is a prerequisite for robust communications in dynamic environments but can simultaneously feed services in health care, human-machine interaction and many other appliances that require context-awareness. 

Next, let us turn our attention to the RSSI at the UE which is minimized (maximized) to transfer the symbol `0' (`1') in BASK. The two green curves in Fig.~\ref{fig5}c evidence that the two RSSI levels are easily distinguishable above the targeted SNR value of 20~dB. Correspondingly, the bit error rate (BER) plotted in Fig.~\ref{fig5}d is very low in this regime. As the SNR is reduced toward 0~dB, the lowest (highest) achievable RSSI increases (decreases), making the two symbols less distinguishable and hence causing the BER to increase. Eventually, below 0~dB, the curves begin to overlap, being dominated by the additive Gaussian noise, and the BER reaches its upper bound of 0.5.

The reasons for the reduced symbol distinguishability at lower SNR values is twofold: on the one hand, the noise increasingly disturbs the desired destructive (constructive) interferences for `0' (`1'); on the other hand, the sensing and localization information based on which these interferences are designed is itself less accurate. To evidence the interplay of localization, sensing and communications more clearly, we plot in Fig.~\ref{fig5}c and Fig.~\ref{fig5}d additionally two benchmarks. The curves in purple correspond to a scenario with perfect context awareness. This benchmark only slightly outperforms the self-adaptive RIS above an SNR of 20~dB because its location and sensing information is only marginally more accurate than our estimates. Below roughly 10~dB in SNR, the difference between the two scenarios grows, quantifying the impact of the imperfectly estimated localization and sensing information on the communications. 

Another important benchmark is that of a context-ignorant RIS (blue). Such a RIS uses two fixed configurations for `0' and `1' that are optimized to yield the best performance on average over many perturber parameter combinations. This strategy makes use of the subset of rays that bounce off RIS elements but not the perturbing object. Clearly, a context-ignorant RIS is expected to perform worse because it fails to control the rays that interact with the uncontrolled perturbers. Indeed, its BER is twice that of the self-adaptive RIS for SNRs above 20~dB. Interestingly, however, below SNRs of 10~dB the context-ignorant RIS performs better because the estimated localization and sensing information of the self-adaptive RIS becomes too imprecise. The benefits of self-adaptive RISs hence unfold only above a threshold-SNR level. This behavior is expected to be general, though the specific threshold-SNR level depends on many parameters: if there were more perturbing objects, fewer rays could avoid all of them, hence the context-ignorant RIS would perform worse.

\section{Challenges, Open Questions, \\and Research Directions}

The above discussions and results provide an initial understanding of the difficulties and potential benefits of operating self-adaptive RISs under the rich-scattering conditions encountered in many realistic 6G deployment scenarios. Important open research questions and challenges remain in order to determine whether RIS-ISAC can help to reach the ambitious goals of 6G:

\textbf{Scalability:} The scalability of the discussed approach to yet more complex scenarios with many more independent perturbers should be investigated. Fundamental limits should be identified on the degree of complexity of channel parametrizations that is learnable with acceptable effort. Naturally, once an individual perturber's impact on the wireless channel is below the noise level, it cannot be sensed efficiently anymore. Moreover, the degree of variability of the perturber parameters may determine the success of the scheme. While the possible combinations of perturber parameters may be clearly defined in a factory setting, an occupied rail car's variations may be too strong. This analysis will hence clarify in which settings a self-adaptive RIS can be deployed and which settings' dynamics are too complex to be suited for a RIS deployment.

\textbf{Motion Forecasting:} The discussed scheme \textit{reacts} to motion by sensing in order to adapt to it. It is enticing to transition to an approach that \textit{anticipates} motion, and hence the evolution of the wireless channels. Given the natural origin of motion (e.g., humans), the use of recurrent neural networks for time-series forecasting is a promising tool to this end. Again, beyond its benefits for communications, these forecasting ``skills'' of a self-adaptive RIS will simultaneously serve other context-aware appliances, too. 

\textbf{Impact of Reverberation Strength:} The reverberation strength in a rich-scattering environment is a double-sided sword for the discussed RIS-ISAC technique. On the one hand, the longer the waves reverberate before being attenuated, the more nonlinear the channel parametrization becomes, making it harder to learn, be it explicitly or implicitly. On the other hand, the longer the waves reverberate, the more sensitive they become to details of the perturbers such that they encode the sought-after location and sensing information more efficiently in the multiplexed field measurements at the auxiliary receivers~\cite{del2021deeply}. Therefore, it is of great importance to understand these trade-offs for 6G RIS-ISAC under rich-scattering conditions.

\textbf{Channel Modelling and Experimentation:} The importance of \textit{physics-based} channel models of \textit{realistic} wireless environments and corresponding \textit{real-life} experimentation is currently underestimated. To reap the full potential of RISs, it must be ensured that developed signal-processing algorithms are compatible with both physics and the targeted deployment scenarios. Accounting for multiple scattering in a rigorous and deterministic manner calls for the development of a new generation of channel models, as well as experimentations outside echo-free test environments.

\section{Conclusion}

If RISs are to be a technological enabler of 6G, they must self-adaptively operate in highly complex settings that can differ remarkably from free space in which RISs are studied to date. In this paper, we have explained why the strongly nonlinear parametrization of rich-scattering wireless channels both through RIS elements and uncontrolled perturbing objects makes full context awareness a prerequisite for RIS-assisted communications. We presented a case study illustrating the resulting interplay of localization and sensing with communications. Our analysis suggests that their convergence in an ISAC fashion is pivotal for the deployment of RISs in commonly discussed 6G settings (e.g., factories) and not merely motivated by limited resources. Looking forward, we discussed key research directions that will determine the feasibility of rich-scattering ISAC and hence RISs in commonly envisioned 6G scenarios.

\bibliographystyle{IEEEtran}
%\bibliography{references}

\begin{thebibliography}{10}
\providecommand{\url}[1]{#1}
\csname url@samestyle\endcsname
\providecommand{\newblock}{\relax}
\providecommand{\bibinfo}[2]{#2}
\providecommand{\BIBentrySTDinterwordspacing}{\spaceskip=0pt\relax}
\providecommand{\BIBentryALTinterwordstretchfactor}{4}
\providecommand{\BIBentryALTinterwordspacing}{\spaceskip=\fontdimen2\font plus
\BIBentryALTinterwordstretchfactor\fontdimen3\font minus
  \fontdimen4\font\relax}
\providecommand{\BIBforeignlanguage}[2]{{%
\expandafter\ifx\csname l@#1\endcsname\relax
\typeout{** WARNING: IEEEtran.bst: No hyphenation pattern has been}%
\typeout{** loaded for the language `#1'. Using the pattern for}%
\typeout{** the default language instead.}%
\else
\language=\csname l@#1\endcsname
\fi
#2}}
\providecommand{\BIBdecl}{\relax}
\BIBdecl

\bibitem{you2021towards}
X.~You \emph{et~al.}, ``Towards {6G} wireless communication networks: Vision,
  enabling technologies, and new paradigm shifts,'' \emph{Sci. China Inf.
  Sci.}, vol.~64, no.~1, pp. 1--74, 2021.

\bibitem{di2019smart}
M.~Di~Renzo \emph{et~al.}, ``Smart radio environments empowered by
  reconfigurable ai meta-surfaces: An idea whose time has come,'' \emph{EURASIP
  J. Wirel. Commun. Netw.}, vol. 2019, no.~1, pp. 1--20, 2019.

\bibitem{alexandropoulos2021hybrid}
G.~C. Alexandropoulos, N.~Shlezinger, I.~Alamzadeh, M.~F. Imani, H.~Zhang, and
  Y.~C. Eldar, ``{Hybrid Reconfigurable Intelligent Metasurfaces: Enabling
  Simultaneous Tunable Reflections and Sensing for 6G Wireless
  Communications},'' \emph{arXiv preprint arXiv:2104.04690}, 2021.

\bibitem{ma2019smart}
Q.~Ma, G.~D. Bai, H.~B. Jing, C.~Yang, L.~Li, and T.~J. Cui, ``Smart
  metasurface with self-adaptively reprogrammable functions,'' \emph{Light Sci.
  Appl.}, vol.~8, no.~1, pp. 1--12, 2019.

\bibitem{ma2020smart}
Q.~Ma, Q.~R. Hong, X.~X. Gao, H.~B. Jing, C.~Liu, G.~D. Bai, Q.~Cheng, and
  T.~J. Cui, ``Smart sensing metasurface with self-defined functions in dual
  polarizations,'' \emph{Nanophotonics}, vol.~9, no.~10, pp. 3271--3278, 2020.

\bibitem{del2019optimally}
P.~del Hougne, M.~Fink, and G.~Lerosey, ``Optimally diverse communication
  channels in disordered environments with tuned randomness,'' \emph{Nat.
  Electron.}, vol.~2, no.~1, pp. 36--41, 2019.

\bibitem{wymeersch2020radio}
H.~Wymeersch, J.~He, B.~Denis, A.~Clemente, and M.~Juntti, ``Radio localization
  and mapping with reconfigurable intelligent surfaces: Challenges,
  opportunities, and research directions,'' \emph{IEEE Veh. Technol. Mag.},
  vol.~15, no.~4, pp. 52--61, 2020.

\bibitem{del2021deeply}
M.~del Hougne, S.~Gigan, and P.~del Hougne, ``{Deeply Subwavelength
  Localization with Reverberation-Coded Aperture},'' \emph{Phys. Rev. Lett.},
  vol. 127, no.~4, p. 043903, 2021.

\bibitem{saigre2022intelligent}
C.~Saigre-Tardif, R.~Faqiri, H.~Zhao, L.~Li, and P.~del Hougne, ``{Intelligent
  meta-imagers: From compressed to learned sensing},'' \emph{Appl. Phys. Rev.},
  vol.~9, no.~1, p. 011314, 2022.

\bibitem{ma2020joint}
D.~Ma, N.~Shlezinger, T.~Huang, Y.~Liu, and Y.~C. Eldar, ``{Joint
  Radar-Communication Strategies for Autonomous Vehicles: Combining Two Key
  Automotive Technologies},'' \emph{IEEE Signal Process. Mag.}, vol.~37, no.~4,
  pp. 85--97, 2020.

\bibitem{de2021convergent}
C.~De~Lima \emph{et~al.}, ``{Convergent Communication, Sensing and Localization
  in 6G Systems: An Overview of Technologies, Opportunities and Challenges},''
  \emph{IEEE Access}, vol.~9, pp. 26\,902--26\,925, 2021.

\bibitem{zhang2022holographic}
H.~Zhang, H.~Zhang, B.~Di, M.~Di~Renzo, Z.~Han, H.~V. Poor, and L.~Song,
  ``{Holographic Integrated Sensing and Communication},'' \emph{IEEE J. Sel.
  Areas Commun.}, 2022.

\bibitem{faqiri2022physfad}
R.~Faqiri, C.~Saigre-Tardif, G.~C. Alexandropoulos, N.~Shlezinger, M.~F. Imani,
  and P.~del Hougne, ``{PhysFad: Physics-Based End-to-End Channel Modeling of
  RIS-Parametrized Environments with Adjustable Fading},'' \emph{arXiv preprint
  arXiv:2202.02673}, 2022.

\bibitem{zhao2020metasurface}
H.~Zhao, Y.~Shuang, M.~Wei, T.~J. Cui, P.~del Hougne, and L.~Li,
  ``Metasurface-assisted massive backscatter wireless communication with
  commodity {Wi-Fi} signals,'' \emph{Nat. Commun.}, vol.~11, no.~1, pp. 1--10,
  2020.

\bibitem{del2020robust}
P.~del Hougne, ``Robust position sensing with wave fingerprints in dynamic
  complex propagation environments,'' \emph{Phys. Rev. Research}, vol.~2,
  no.~4, p. 043224, 2020.

\end{thebibliography}


\end{document}